# Gender Dynamics in Software Engineering: Insights from Research on Concurrency Bug Reproduction

Tarannum Shaila Zaman[1], Macharla Hemanth Kishan[2], Lutfun Nahar Lota[3]

[1] University of Maryland Baltimore County, Baltimore, MD, USA

[2] SUNY Polytechnic Institute, Utica, NY, USA

[3] Islamic University of Technology, Gazipur, Bangladesh

Email: zamant@umbc.edu, macharh@sunypoly.edu, lota@iut-dhaka.edu

*Abstract*—Reproducing concurrency bugs is a complex task due to their unpredictable behavior. Researchers, regardless of gender, are contributing to automating this complex task to aid software developers. While some studies have investigated gender roles in the broader software industry, limited research exists on gender representation specifically among researchers working in concurrent bug reproduction. To address this gap, in this paper, we present a literature review to assess the gender ratio in this field. We also explore potential variations in technique selection and bug-type focus across genders. Our findings indicate that female researchers are underrepresented compared to their male counterparts in this area, with a current male-to-female author ratio of 29:6. Through this study, we emphasize the importance of fostering gender equity in software engineering research, ensuring a diversity of perspectives in the development of automated bug reproduction tools.

## I. Introduction

In every stage of software engineering, humans play a central role. This field is highly human-centered, both in academia and industry. However, gender representation in software engineering is disproportionate [1]. Although it is wellknown that the proportion of female researchers in software engineering is relatively low, our goal is to identify specific statistics to understand the gender dynamics in the area of reproducing concurrency bugs. Reproducing these bugs is an essential aspect of debugging concurrent programs [2].

Bug reproduction is the first and one of the important phases of software debugging [3]. Generally, debugging is a diagnosis procedure to determine whether the program meets the expectations or not. Diagnostic information may come from bug reports [4] generated during testing or from various logs, such as console logs [2], network logs [5], and system logs [6]. It is necessary to reproduce the bug to be confirmed and to observe the behavior of the bug. Bug reproduction also enables further analysis of the issue. Once the bug is reproduced, developers investigate its root cause, localize it, and then devise a solution to resolve it [7]. A recent study at Google [8] emphasizes that reproducing the bug is the essential first step toward solving it. However, reproducing concurrency bugs is particularly challenging due to their non-deterministic nature [2], [6]. In a survey at Microsoft [9], more than 70% of developers said that reproducing concurrency bugs is very difficult.

In this paper, we focus on the gender dynamics among researchers working on automated bug reproduction techniques, particularly those addressing concurrency bugs. Researchers have done a series of works to detect and reproduce concurrency bugs. Over years of research in concurrency, we have observed a noticeable lack of gender diversity in this area. This observation motivated us to conduct a statistical analysis of gender representation through literature reviews focused on reproducing concurrency bugs. In future, we aim to expand this analysis to other areas of debugging, such as fault localization and bug resolution [2], [6].

In this literature review, we focus on research that aims to automate the reproduction of concurrency bugs, but we believe it also reveals trends in gender diversity in debugging these types of issues. Diversity is important in research, covering not only gender but also culture, religion, and location. However, we focus only on gender diversity in this paper. Our goal is to show the current state of gender representation in one area (Reproducing Concurrent Bugs) of software engineering, which may give us a sense of what to expect in other areas. We also look at whether there's any link between gender and the choice of methods or types of bugs studied. We hope this study sheds light on gender diversity among researchers working on concurrency bugs and encourages steps toward gender equity in this field and beyond. This review paper also aims to encourage more discussion on gender issues and ways to improve diversity and equity. The following research questions are addressed in this paper.

RQ1: What is the ratio of male to female researchers working in the field of reproducing concurrency bugs? RQ2: Is there any role of different genders in analyzing any specific types of bugs?

RQ3: What is the role of different genders in leveraging different techniques to reproduce concurrent bugs?

## II. Motivation and Related Works

In this section, we present our motivation and the difference of our work with the existing related works.

Motivation: Reproducing and diagnosing concurrency failures in deployed systems is challenging due to limited production data [10], [6] and the non-deterministic nature of these bugs [11]. A Microsoft survey found that over 70% of developers struggle with reproducing concurrency bugs [9], leading to extensive research on automation. However, no comprehensive review focuses solely on this area, nor do existing studies examine gender representation in the field.

We investigate gender roles in concurrency bug reproduction research by analyzing author demographics. Identifying participation trends across genders may help encourage more women to enter the field and promote diversity in software engineering research.

Related Works: There are some existing literature reviews relevant to our work [12], [13], [14]. Table I provides an overview of these studies, including publication year, focus area, and differences with this work. While none of these works focus precisely on our research area, they address related topics. For example, Wang et al. [12] conduct a systematic review centered on techniques for automatically detecting, exposing, avoiding, and fixing concurrency bugs, yet they do not cover automated bug reproduction techniques. Asadollah et al. [13] review a decade of research (2005–2014) on debugging concurrent and multicore software, encompassing tasks beyond bug reproduction. Wambugu et al. [14] present a brief review on fault localization, program slicing, and delta debugging, examining tools for automating debugging but not specifically addressing concurrent bugs.

| Cite | Year | Focused Area | Difference |
|------|------|--------------|------------|
| [12] | 2018 | automated concurrency bug detection, exposing, avoidance, and fixing | no focus on concurrency bug reproduction, and on the role of different genders |
| [13] | 2017 | debugging concurrent and multicore software | no survey on the role of different genders |
| [14] | 2017 | discusses fault localization, program slicing, and delta debugging techniques | no focus on concurrent programs, bug reproduction, and no survey on the role of different genders |

TABLE I Existing literature reviews related to our work

Notably, none of these works analyze gender representation or roles within their research focus. Understanding the gender distribution and roles in a research area is essential for fostering an unbiased environment. In this paper, we present an analysis of gender roles in the field of concurrent bug reproduction research.

## III. Background

In this section, we define key terminologies and outline the scope of our analysis. Throughout the remainder of the paper, these terms will be used to explain the findings of this survey.

### A. Concurrent Systems

A concurrent system executes multiple processes or threads that interact simultaneously [15]–[17]. These run on the same or different machines, depending on the system's architecture [18].

Concurrent systems fall into two categories: multi-threaded systems, where threads share memory within a process, and multi-process systems, where each process has a separate address space [10], [19]. Concurrency bugs appear as intraprocess bugs in multi-threaded systems and inter-process bugs in multi-process systems, with the latter posing greater communication challenges [10].

### B. Reproducing Concurrency Bugs

We categorize existing research on reproducing concurrency bugs into two groups: thread-level and process-level concurrency bug reproduction.

Thread-Level Concurrency Bug Reproduction: Many studies focus on detecting or reproducing intra-process concurrency bugs, as shown in the first column Table II. These include non-deadlocking bugs like atomicity violations, order violations, and data races, which are frequent and challenging to manage [20], [21]. Deadlocks, which can halt a system unpredictably, have led to various detection techniques [22], [23].

Process-Level Concurrency Bug Reproduction: Table III highlights research on inter-process concurrency bugs, referred to as process-level bugs in this survey. Dynamic analysis techniques are commonly used to reproduce these bugs.

## IV. Survey Design

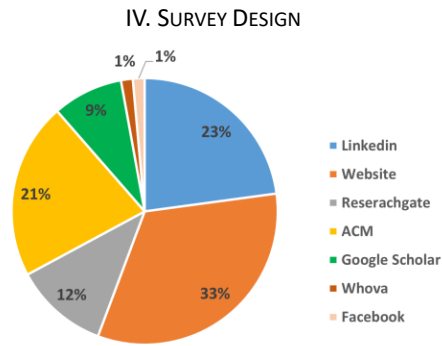

Fig. 1. Sources used to identify gender of the authors

The second and third authors of this paper individually search on *Google Scholar* using two different search strings: i) "concurrent bug reproduce" and ii) "concurrent bug regeneration." These searches yield a total of 27,900 and 11,500 research papers, respectively, as of October 2024. We define inclusion and exclusion criteria for selecting the final set of papers as follows:

Inclusion Criteria (IC): i) Studies focus on software bug reproduction. ii) Studies were published online between 2000 and 2024. iii) Studies specifically address concurrent bugs.
Exclusion Criteria (EC): i) Studies unrelated to concurrent

programming. ii) Studies not focused on bug reproduction. iii) Studies not written in English.

Based on these criteria, the second and third authors independently select papers. To assess the consistency of their selections, Cohen's Kappa value ($K > 0.8$), as proposed by Perez et al. [24], is calculated. Ultimately, they select 22 papers. The first author reviews all the final selections to ensure that they meet the inclusion criteria.

After selecting the final 22 papers, we searched for all the authors to identify their gender. For gender identification, we reviewed the authors' photos and any pronouns mentioned in their public profiles. In total, there were 79 authors across the 22 papers. We conducted online searches to determine their gender. Figure 1 illustrates the sources used for gender identification. According to the pie chart, 33% of the gender information was gathered from *Personal Websites*, 23% from *LinkedIn*, 21% from *ACM researcher profiles*, 12% from *ResearchGate*, 9% from *Google Scholar*, and 1% each from *Whova* and *Facebook*.

## V. RESULT ANALYSIS

| Name | Year | Type of bug | Techniques | Authors |
|---|---|---|---|---|
| Pres [25] | 2009 | Non-deadlocking bugs | Record and replay | T:7, M:3, F:3, N:1 |
| DEADLOCK FUZZER [26] | 2009 | Deadlocking Bugs | Dynamic analysis | T:4, M:3, F:1 |
| Leap [27] | 2010 | Non-deadlocking bugs in Java Program | Record and replay | T:3, M:2, F:0, N:1 |
| Stride [28] | 2012 | Non-deadlocking bugs in Java Program | Record and replay | T:3, M:2, F:0, N:1 |
| CoopREP [29] | 2012 | Non-deadlocking bugs | Record and replay | T:3, M:3, F:0, N:0 |
| DynaRace [30] | 2012 | TOCTTOU Races | Dynamic analysis | T:2, M:2, F:0, N:0 |
| CONCURRIT [31] | 2013 | Non-deadlocking bugs | Domain-specific language | T:4, M:4, F:0, N:0 |
| CARE [32] | 2014 | Non-deadlocking bugs in long-running programs | Record and replay | T:5, M:5, F:0, N:0 |
| ReproLite [33] | 2014 | Concurrent Bugs in Cloud Systems | Domain-specific language | T:4, M:2, F:1, N:1 |
| Deadlock detection and reproduction [34] | 2014 | Deadlocking Bugs | Dynamic analysis and Record replay | T:2, M:1, F:1, N:0 |
| ConCrash [35] | 2017 | Non-deadlocking bugs | Search, pruning, and exploring interleavings | T:3, M:2, F:0, N:1 |
| Concurrency race debugging [36] | 2018 | Data-races | Dynamic Slicing | T:6, M:5, F:0, N:1 |
| RAProducer [37] | 2021 | Data -Races | Static analysis on binaries | T:6, M:3, F:1, N:2 |
| ThreadRadar [38] | 2021 | Thread-level Races | statistical profiling, clustering | T:3, M:3 |
| Hippodrome [39] | 2023 | Data- Races | Static Analysis | T:6, M:4, F:1, N:1 |
| Causal Debug [40] | 2024 | Thread-level Races | Roll-back and Replay | T:2, M:2 |

TABLE II
SUMMARY OF RESEARCH WORKS CONDUCTED FOR REPRODUCING THREAD-LEVEL CONCURRENCY BUGS

*Techniques* = Used techniques to solve the problem, *Authors* = Show the details of the authors: T: Total number of authors, M: Total number of Male authors, F: Total number of Female authors, N: Total number of authors whose gender can not be identified.

In this section, we provide the answers to the research questions. Tables II, and III summarize the survey results for thread-level concurrency bugs and process-level concurrency bugs respectively. Column 1 of each table lists the research paper name and reference, Column 2 indicates the publication year, Column 3 identifies the types of bugs targeted by the respective research, Column 4 outlines the techniques used to address the problem, and the last column presents the number of authors along with their gender for each research work.

RQ1: Ratio of male to female researchers working in the field of reproducing concurrency bugs:

Column 3 of Table II and Table III presents the total number of authors (denoted as T), male authors (M), and female authors (F) for each paper. *N* indicates the number of authors whose gender could not be identified through online searches. Figure 2 illustrates the gender distribution of authors in research on reproducing thread-level concurrency bugs.

| Name | Year | Type of bug | Techniques | Genders |
|---|---|---|---|---|
| ODR [41] | 2009 | Non-deadlocking bugs | Record and replay | T:2, M:2, F:0, N:0 |
| Concurrency bug reproduction. [42] | 2010 | Heisen bugs | Core dump analysis and execution indexing | T:3, M:3, F:0, N:0 |
| CLAP [43] | 2013 | Non-deadlocking bugs | Dynamic analysis | T:3, M:3, F:0, N:0 |
| SimRacer [44] | 2013 | System-level races | Dynamic analysis | T:3, M:2, F:1, N:0 |
| DESCRY [10] | 2017 | System-level races | Dynamic analysis, static analysis and symbolic execution | T:3, M:1, F:2, N:0 |
| RedPro [45] | 2022 | System-level races | Dynamic Binary Analysis | T:2, M:1, F:1 |

TABLE III
SUMMARY OF RESEARCH WORKS CONDUCTED FOR REPRODUCING PROCESS-LEVEL CONCURRENCY BUGS

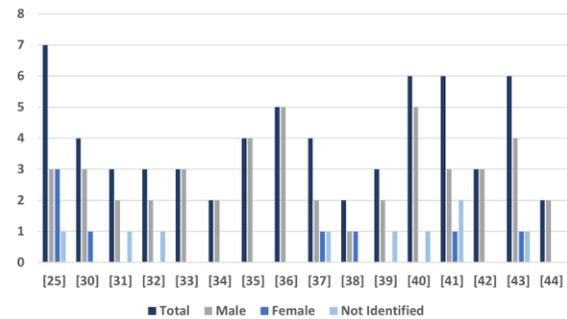

Fig. 2. Gender dist. of authors in research on reproducing thread-level bugs

According to Figure 2, five papers [29], [30], [31], [32], [38] are authored exclusively by males. The papers *Pres* [25] and *Trace-driven dynamic deadlock detection and reproduction* [34] have an equal number of male and female authors, though one author's gender remains unidentified in *Pres* [25]. In all other papers, including [26], [27], [28], [33], [37], [39] the number of male authors exceeds that of female authors. In total, 63 authors contributed to these 16 papers, with 46 identified as male, 8 as female, and 9 whose gender could not be determined. The ratio of male to female authors is 23:8.

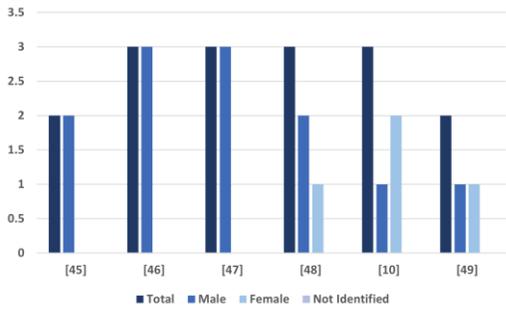

Fig. 3. Gender dist. of authors in research on reproducing process-level bugs

Figure 3 shows that all authors in three papers [41], [42], [43] focusing on system-level concurrency bugs are male. The paper Redpro [45] has an equal number of male and female authors, Descry [10] has a higher number of female authors, while SimRacer [44] has more male authors. In total, 16 authors contributed to these six papers, with 12 identified as male and 4 as female, resulting in a male-to-female ratio of 3:1.

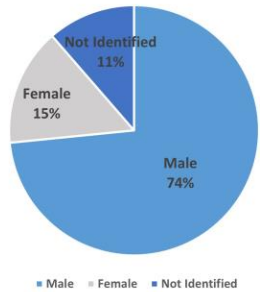

Fig. 4. Percentage of Different Genders

In total, there are 79 authors across all 22 papers, with 58 identified as male, 12 as female, and 9 whose gender remains unknown. Figure 4 illustrates the overall distribution: 74% of the authors are male, 15% are female, and the gender of 11% could not be identified through online sources. The overall ratio of male to female authors is 29:6.

RQ2: Role of different genders in analyzing different types of bugs: Column 3 of Table II and Table III presents the specific types of targeted bugs in multi-threaded and multi-process concurrent applications. In this research question, we explore the relationship between the types of bugs and the gender of the authors. Figure 5 represents the total number of male (M), female (F), and not identified gender (N) authors for six different types of concurrent bugs.

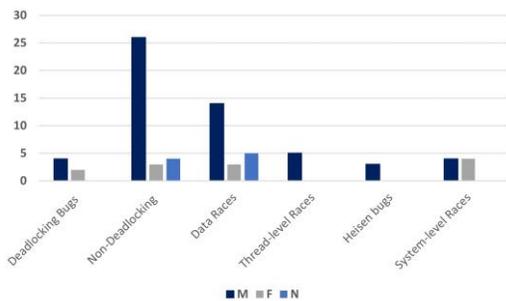

|   | DB | NonDB | DataR | T-level | HeisenB | Sys-level |
|---|---|---|---|---|---|---|
| M | 66% | 79% | 63% | 100% | 100% | 50% |
| F | 33% | 9% | 13% | 0% | 0% | 50% |
| N | 0% | 12% | 23% | 0% | 0% | 0% |

TABLE IV
PCT. OF DIFFERENT GENDERS WORKING ON VARIOUS TYPES OF BUGS

DB: Deadlocking Bugs, NonDB: Non-deadlocking bugs, DataR: Data Races, T-level: Thread-level races, HeisenB: Heisen Bug, Sys-level: System-level Races.

Fig. 5. No. of authors of different genders working on various types of bugs

There are 2 deadlocking bugs, 9 non-deadlocking bugs, 4 data races, 2 thread-level races, 1 Heisenbug, and 3 systemlevel races. All authors working on reproducing thread-level races and Heisenbugs are male. An equal number of male and female authors work on reproducing system-level races. Rows 1, 2, and 3 of Table IV present the percentage of male, female, and unidentified authors for each type of bug. The data indicates that, except for system-level races, the percentage of male researchers is higher than that of female researchers across all bug types.

RQ3: Role of different genders in leveraging techniques to reproduce concurrent bugs: Column 4 of table II, and table III presents the various techniques employed by researchers to address the challenge of automatically reproducing concurrent

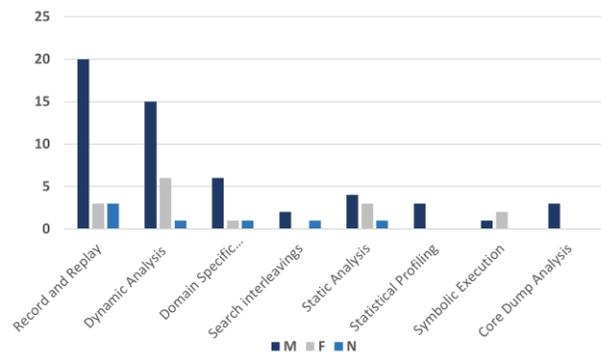

Fig. 6. No. of authors of different genders working on various techniques

bugs in multi-threaded and multi-process applications. A total of 8 different techniques are utilized for this research problem. Among these, the Record and Replay technique is used in 7 studies, Dynamic Analysis in 8 studies, Domain-Specific Language in 2 studies, and Search Interleavings in 1 study. Additionally, Static Analysis, Statistical Profiling, Symbolic Execution, and Core Dump Analysis are each used in 3 different studies. Figure 6 illustrates the total number of male (M), female (F), and not-identified (N) researchers across 22 different studies. Rows 1, 2, and 3 of Table V displays the percentage of male (M), female female (F), and not-identified

| | RR | DA | DSL | SI | SA | SP | SE | CDA |
|---|---|---|---|---|---|---|---|---|
| M | 77% | 68% | 75% | 67% | 50% | 100% | 33% | 100% |
| F | 12% | 27% | 13% | 0% | 38% | 0% | 67% | 0% |
| N | 12% | 5% | 12% | 33% | 13% | 0% | 0% | 0% |

TABLE V
PCT. OF DIFFERENT GENDERS WORKING ON VARIOUS TECHNIQUES

RR: Record and Replay, DA: Dynamic Analysis, DSL: Domain Specific Language, SL: Search interleavings, SA: Static Analysis, SP: Statistical Profiling, SE: Symbolic Execution, CDA: Core Dump Analysis

(N) researchers across 22 different studies. Rows 1, 2, and 3 of Table V display the percentage of male (M), female (F), and not-identified researchers for each of the 8 different techniques. According to the table, the Statistical Profiling and Core Dump Analysis techniques are exclusively used by male researchers. Symbolic Execution is the only technique where female researchers outnumber male researchers. For the remaining five techniques, the percentage of male authors is higher than that of female authors.

## VI. DISCUSSION AND FUTURE SCOPE

Based on our analysis in Section V, female researcher participation in reproducing concurrency bugs remains low. Given the importance of concurrency bugs in software engineering, achieving greater gender equity in this area is desirable. While this review highlights recent gender trends, it offers only a partial view, as it focuses on a single subfield of software debugging. In the future, we plan to expand our survey to cover the entire field of software debugging.

We also aim to explore challenges faced by female researchers in this domain. To assess whether gender identity influences specific difficulties, we plan to survey researchers of various genders. Understanding factors behind lower participation can help develop strategies for broader engagement. Additionally, through researcher interviews, we seek to identify potential gender biases in academia and industry, ultimately promoting greater inclusivity in this field.

## VII. THREATS TO VALIDITY

Three types of threats to validity are considered to confirm the validity of this work.

Threats to internal Validity: A key threat to internal validity is selection bias in our criteria. To minimize bias toward specific authors, venues, or tools, we use a keyword search that excludes these factors. Our second and third authors independently conduct this search, guided by standardized inclusion and exclusion criteria (Section IV). The first author then verifies that all selected papers meet these criteria.

Another threat is accurately identifying an author's gender. We mitigate this by reviewing public profiles and photos. This study classifies gender as male or female, with other identities not considered. If gender cannot be determined, it is labeled as "not identified."

Threats to External Validity: The main threat to external validity is whether our selection criteria adequately represent the field. To mitigate this, we conducted a keyword search and identified all mutually related works. Another concern is the accuracy of gender identification based on author photos. To reduce bias, we verify information across multiple sources, including LinkedIn, personal websites, and ResearchGate.

Threats to Conclusion Validity: To reduce the threats to the conclusion validity of this work, we kept the date and time of the web searches. As the content of the internet is changing continuously, it cannot be guaranteed that a later search with the same keywords will yield the same results as ours.

## VIII. CONCLUSION

Reproducing concurrency bugs is a complex and resourceintensive task, requiring substantial manpower, time, and financial investment. To simplify and automate this process, researchers have made significant strides. However, there is currently no study that provides insights into the gender representation of the researchers within the field of concurrency bug reproduction. Understanding the roles and statistics of various genders in software engineering research could reveal potential patterns in technique selection or problem types approached by different genders. Our aim is to promote gender equity across all areas of software engineering. In this paper, we present data on gender participation specifically in the field of reproducing concurrency bugs. We hope these insights will inspire concrete actions toward achieving gender equity in software engineering.

## IX. ACKNOWLEDGMENT

This work was supported in part by NSF grants CCF2348277 and CCF-2518445.